\documentclass[aip,apl,reprint,superscriptaddress,citeautoscript]{revtex4-1}
\usepackage{graphicx,color,amsmath,amssymb,bm,dcolumn}
\usepackage{soul}

\begin{document}
\title{Room-temperature stability of excitons and transverse-electric polarized deep-ultraviolet luminescence in atomically thin GaN quantum wells}

\author{Dylan Bayerl}
\affiliation{Department of Materials Science and Engineering, University of Michigan, Ann Arbor, MI 48109, USA}
\author{Emmanouil Kioupakis}
\email{kioup@umich.edu}
\affiliation{Department of Materials Science and Engineering, University of Michigan, Ann Arbor, MI 48109, USA}

\date{\today}

\begin{abstract}
We apply first-principles calculations to study the effects of extreme quantum confinement on the electronic, excitonic, and radiative properties of atomically thin (1 to 4 atomic monolayers) GaN quantum wells embedded in AlN.
We determine the quasiparticle band gaps, exciton energies and wave functions, radiative lifetimes, and Mott critical densities as a function of well and barrier thickness.
Our results show that quantum confinement in GaN monolayers increases the band gap up to 5.44 eV and the exciton binding energy up 215 meV, indicating the thermal stability of excitons at room temperature.
Exciton radiative lifetimes range from 1 to 3 ns at room temperature, while the Mott critical density for exciton dissociation is approximately $10^{13}$ cm$^{-2}$.
The luminescence is transverse-electric polarized, which facilitates light extraction from c-plane heterostructures. 
We also introduce a simple approximate model for calculating the exciton radiative lifetime based on the free-carrier bimolecular radiative recombination coefficient and the exciton radius, which agrees well with our results obtained with the Bethe-Salpeter equation predictions.
Our results demonstrate that atomically thin GaN quantum wells exhibit stable excitons at room temperature for potential applications in efficient light emitters in the deep ultraviolet, as well as room-temperature excitonic devices.
 \end{abstract}

\maketitle

\begin{figure}
\includegraphics[scale=0.55]{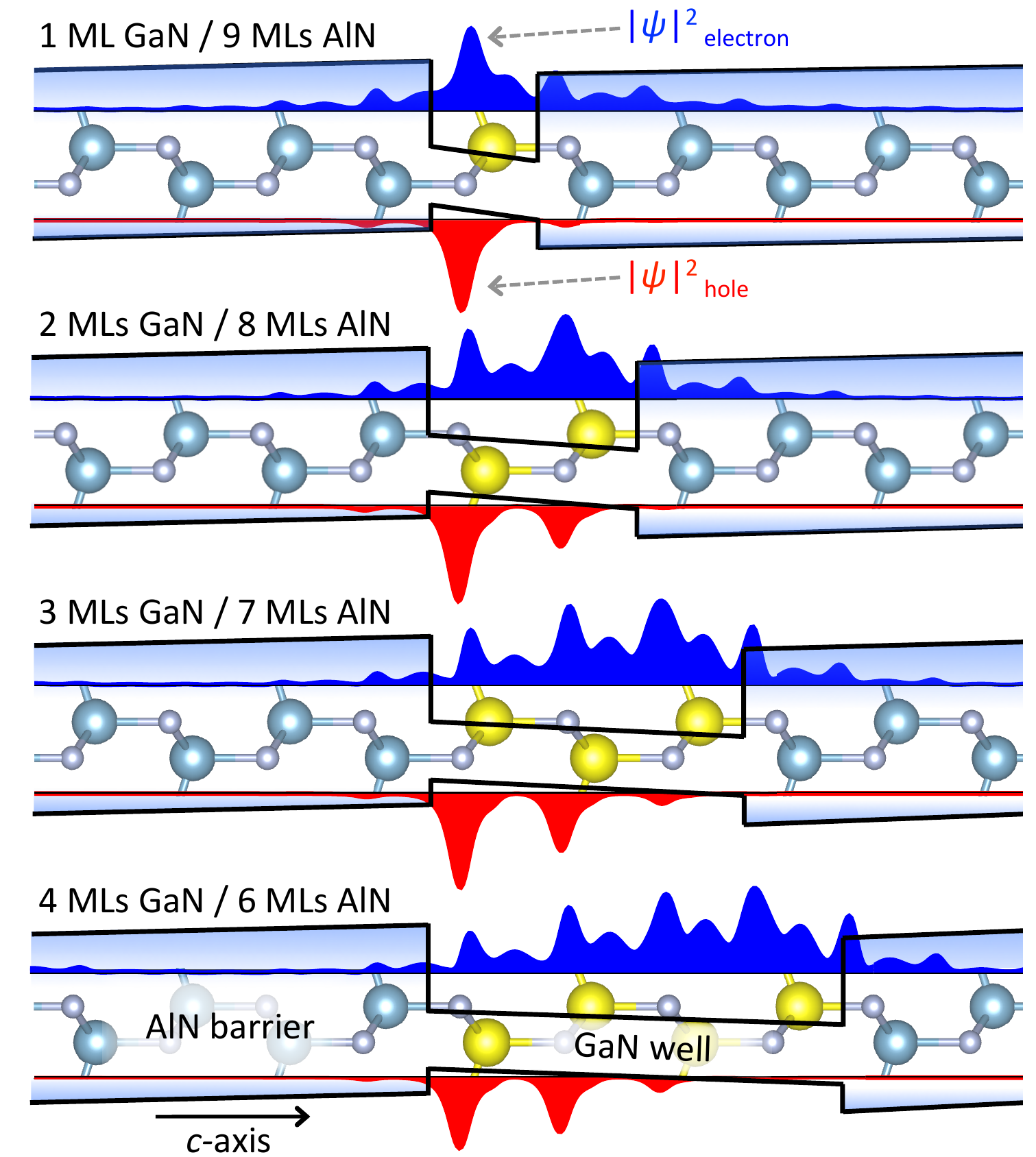}
\caption{\label{fig:structure}
(Color online) Crystal structures, band diagrams, and carrier wave functions for a series of atomically thin GaN QWs in AlN barriers
with well thickness ranging from 1 ML (top) to 4 MLs (bottom).
The wave functions of electrons at the minimum conduction (blue) and holes at the top valence (red) bands at $\Gamma$
reveal strong carrier localization within the GaN well (yellow cations).
The spatial overlap of the carrier wave functions decreases with increasing well thickness due to the strong electrical polarization perpendicular to the layers
and modulates the carrier recombination properties.
}
\end{figure}

The group-III nitrides are established materials for solid-state lighting and high-power electronics.
Recently, attention has focused on AlGaN alloys for light emission in the deep ultraviolet (UV) for applications in germicidal sterilization, water purification, gas sensing, and UV curing.\cite{kneissl2016brief}
However, challenges relating to p-type doping\cite{Pandey2019} and light extraction need to be overcome to increase the efficiency of AlGaN-based UV light-emitting diodes (LEDs).
An alternative approach uses atomically thin binary GaN quantum wells (QWs)\cite{taniyasu2011polarization,kamiya2011structural} with potentially improved deep-UV light emission and extraction efficiencies.
Both theory\cite{bayerl2016deep} and experiment\cite{Verma2013,Selles2016,Sarwar2016} demonstrated that the strong quantum confinement in atomically thin GaN shifts the emission wavelength into the deep UV, and several groups reported high efficiency deep-UV emission\cite{Rong2016,islam2017deep,Haughn2019,Kobayashi2019} with transverse electric (TE) polarization\cite{Zhang2011, Rong2016,Liu2018} in LEDs with atomically thin GaN or AlGaN QWs.
Such atomically thin semiconductors resemble two-dimensional materials such as graphene and transition-metal dichalcogenides, which have recently risen to prominence for potential applications in atomically thin electronics and quantum devices.
Atomically thin nitrides are therefore a promising avenue to realize the novel functional properties of two-dimensional materials on an established semiconductor platform.

In this work, we employ density functional and many-body perturbation theory to investigate the fundamental electronic, excitonic, and radiative  properties of atomically thin GaN QWs in AlN barriers.
We demonstrate that these heterostructures have band gaps in the deep UV and emit TE ($\vec{E}\perp c$) polarized light (similar to bulk GaN) for higher light-extraction efficiency compared to AlGaN. The strong Coulomb interaction in the atomically thin regime stabilizes excitons at room temperature and results in shorter radiative lifetimes than free-carrier recombination.
Our results demonstrate the advantages of atomically thin QWs for efficient light emission and the realization of room-temperature excitonic devices on an established semiconductor platform.

Our first-principles methodology is based on density functional theory and many-body perturbation theory,
which accurately predict electronic, excitonic, and optical properties of materials.\cite{Rondinelli2015}
We calculated the electronic band structures with the GW method and excitonic properties with the Bethe-Salpeter equation (BSE) method using the Quantum ESPRESSO\cite{quantumespresso}, BerkeleyGW\cite{berkeleyGW}, and Wannier90\cite{mostofi2008wannier90} codes (details in \color{black}the Supplementary Material\color{black}).
We determined the bimolecular radiative recombination coefficients of free carriers using the method of Kioupakis \emph{et al},\cite{kioupakis2013temperature,mengle2016first} and the radiative lifetimes of excitons from BSE calculations as in Palummo \emph{et al}.\cite{palummo2015exciton}

We investigated atomically thin GaN QWs between 1 monolayer (ML) and 4 MLs thick separated by AlN barriers between 1 ML and 9 MLs thick. 
To simulate pseudomorphic growth on AlN, the basal plane lattice constant was fixed to that of bulk AlN (0.3112 nm)\cite{levinshtein2001properties} while relaxing the atom positions and c-axis length.
A representative subset of these structures is shown in Fig.~\ref{fig:structure}, along with the plane-averaged electron and hole wave functions.

\begin{figure}
\includegraphics[scale=0.6]{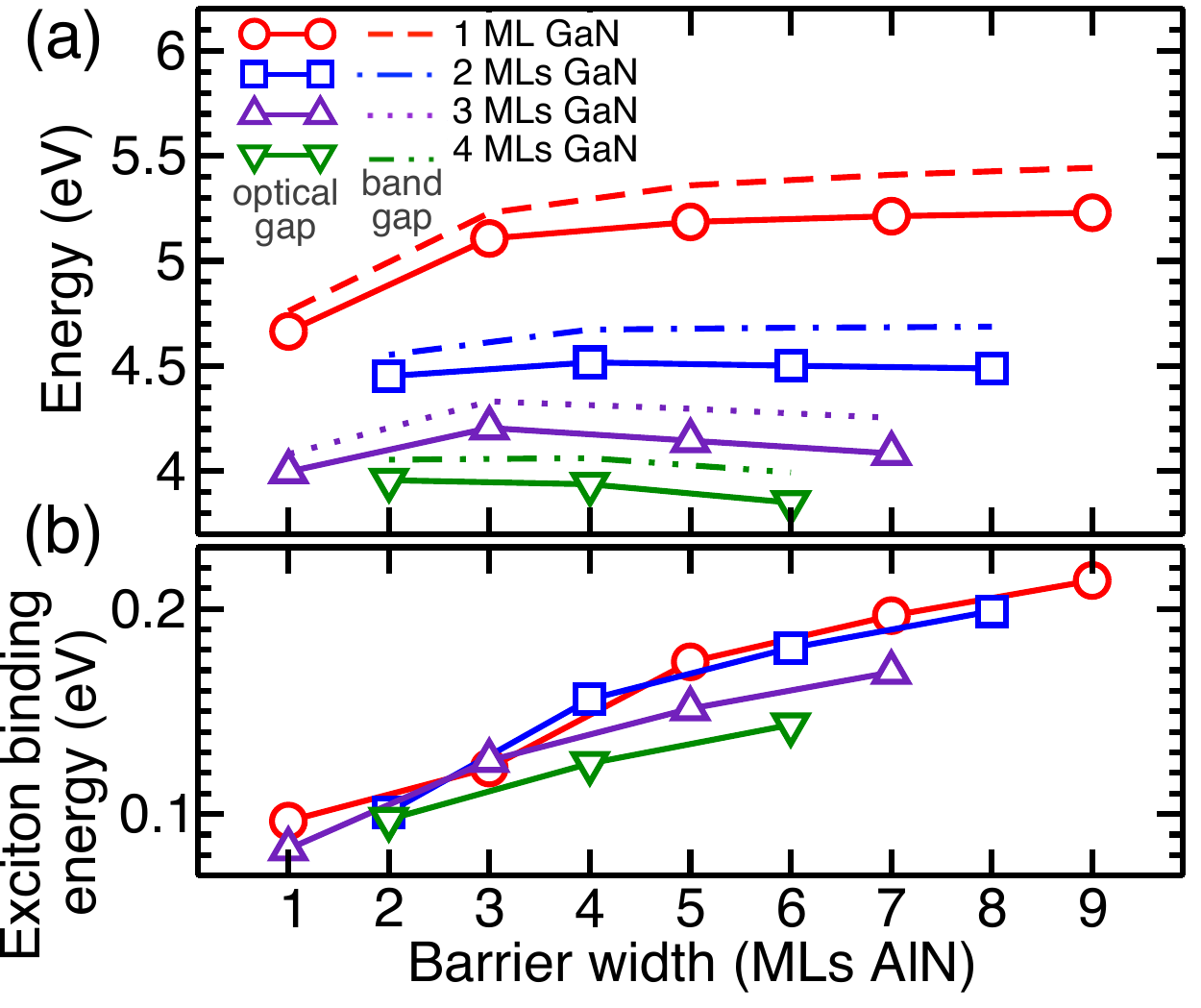}
\caption{\label{fig:elec_opt_gaps}
(Color online) (a) Band gap (dashed lines), optical gap (\emph{i.e.}, lowest singlet exciton energy, solid lines), and 
(b) lowest singlet exciton binding energy of atomically thin GaN QWs in AlN barriers 
as a function of the well and barrier thickness.
The luminescence energy ranges from 3.8 to 5.5 eV and the exciton binding energy from 83 to 215 meV. 
}
\end{figure}

\begin{figure}
\includegraphics[scale=0.7]{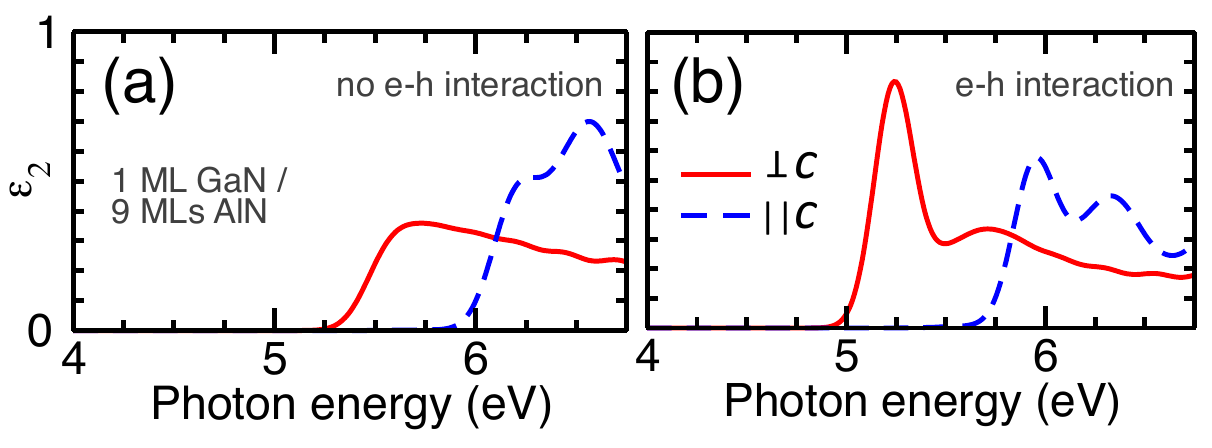}
\caption{\label{fig:epsilon2}
\color{black}(Color online) Imaginary part of the dielectric function $\epsilon_2$ of a short-period GaN/AlN superlattice both (a) excluding and (b) including electron-hole interactions. The absorption/emission onset occurs at lower energies for light polarized perpendicular to the $c$ axis (solid lines) than parallel to it (dashed lines) leading to TE-polarized emission.\color{black}
}
\end{figure}

Our calculated electronic band gaps and optical gaps (i.e., lowest singlet exciton energies) are shown in Fig.~\ref{fig:elec_opt_gaps}a and listed in Table~\ref{tab:ultrathin_properties}. The full band structures for the 1-4 ML simulated structures with the thickest barriers are provided in the supplementary material.
The band gap is direct at $\Gamma$ and corresponds to electron and hole states localized in the GaN QW (Fig. \ref{fig:structure}). 
The gap increases to higher values than bulk GaN (3.4 eV) due to quantum confinement in the atomically thin wells rather than alloying with AlN. 
Moreover, the valence-band structure is GaN-like, i.e., the top two valence bands are the heavy and light hole bands, while the crystal-field band is shifted lower in energy by confinement due to its lower effective mass along the c-axis direction. This valence-band ordering ensures that the emitted light is TE-polarized \color{black}(Fig. \ref{fig:epsilon2})\color{black}, which enables higher photon extraction efficiency in c-plane LEDs, as also reported experimentally.\cite{Rong2016,Liu2017}
For 1 ML and 2 ML GaN QWs the band gap increases with increasing barrier thickness up to values of 5.44 eV and 4.69 eV, respectively, in excellent agreement with experimental results.\cite{bayerl2016deep}
For 3 ML and 4 ML GaN QWs with more than 3 MLs AlN barrier, the gap decreases with increasing barrier thickness due to the quantum-confined Stark effect (QCSE):
thicker wells and thicker barriers increase the polarization field in the QW and further spatially separate electrons and holes along the potential gradient (Fig. \ref{fig:structure}).
The QCSE simultaneously reduces the band gap and electron-hole wave function overlap.
The electron-hole overlap integral, $|F|^2 = [\int \psi_e(z) \psi_h(z) dz]^2$, where $\psi_e(z)$ and $\psi_h(z)$ are the plane-averaged electron and hole envelope functions, respectively, decreases from 0.88 to 0.55 as QW thickness increases from 1 to 4 MLs (Table~\ref{tab:ultrathin_properties}).
Note the QCSE has little influence on the 1 ML and 2 ML QWs, since the extreme confinement in atomically thin wells prevents electrons and holes from spatially separating.


We also determined the exciton binding energy from the difference between the calculated band gap and lowest singlet exciton energy (Fig. \ref{fig:elec_opt_gaps}b and Table~\ref{tab:ultrathin_properties}).
The binding energy increases with decreasing QW or increasing barrier thickness due to the strong confinement of electrons and holes in thinner wells or by thicker barriers.
The most confining structure with a 1 ML GaN QW and 9 ML AlN barrier exhibits an exciton binding energy of 215 meV, which exceeds the binding energy of bulk GaN (20 meV)\cite{Muth1997} and the thermal energy ($k_b T$, where $k_b$ is Boltzmann's constant) at room temperature (26 meV) by approximately one order of magnitude. Such large exciton binding energies have also been estimated with model calculations.\cite{Derevyanchuk2018} The binding energy of these strongly confined excitons also exceeds the value (4 times the bulk binding energy) determined by Bastard \emph{et al}. in the two-dimensional limit.\cite{Bastard1982} One reason for the discrepancy is that the analysis by Bastard \emph{et al}. assumes the same dielectric constant both for the wells and the barriers. However, the dielectric constant of AlN is smaller than GaN due to its wider gap, which reduces the screening of the Coulomb attraction between electrons and holes, and strengthens their binding. Another reason is the reduced screening by ions in the two-dimensional limit. Exciton binding energies in bulk semiconductors such as GaN are of the order of 10 meV and are affected by ionic screening of the Coulomb interaction: polar phonons in materials with light elements such as oxides and nitrides have frequencies on the order of 100 meV (92 meV for GaN\cite{Davydov1998}), thus ions can move fast enough to screen the Coulomb attraction between electrons and holes. This is also the case for SnO$_2$, in which the calculated exciton binding energies are in much better agreement with experiment if the static dielectric constants are employed (which account for both electronic and ionic screening) rather than the high-frequency values (electronic screening only).\cite{Schleife2011}
In atomically thin GaN, however, the characteristic energy of excitons (215 meV) is higher than the phonon frequencies and thus the Coulomb attraction is primarily screened by other electrons only.
Both of these factors amplify the electron-hole Coulomb attraction in atomically thin GaN QWs, which stabilize excitons at room temperature and may lead to the formation of higher order quasiparticles such as unconventional biexcitons\cite{Honig2014} on an established semiconductor platform.

We also evaluated the exchange splitting between singlet and triplet exciton states, $\Delta E_{ST}$, which is also increased by quantum confinement. Our results (Table~\ref{tab:ultrathin_properties}) show that even the highest value of 21 meV, which occurs for the thinnest (1 ML) GaN QWs, is comparable to $k_bT$ at room temperature and thus thermal fluctuations can cause transitions to the radiative singlet state. 

\begin{figure}
\includegraphics[scale=0.6]{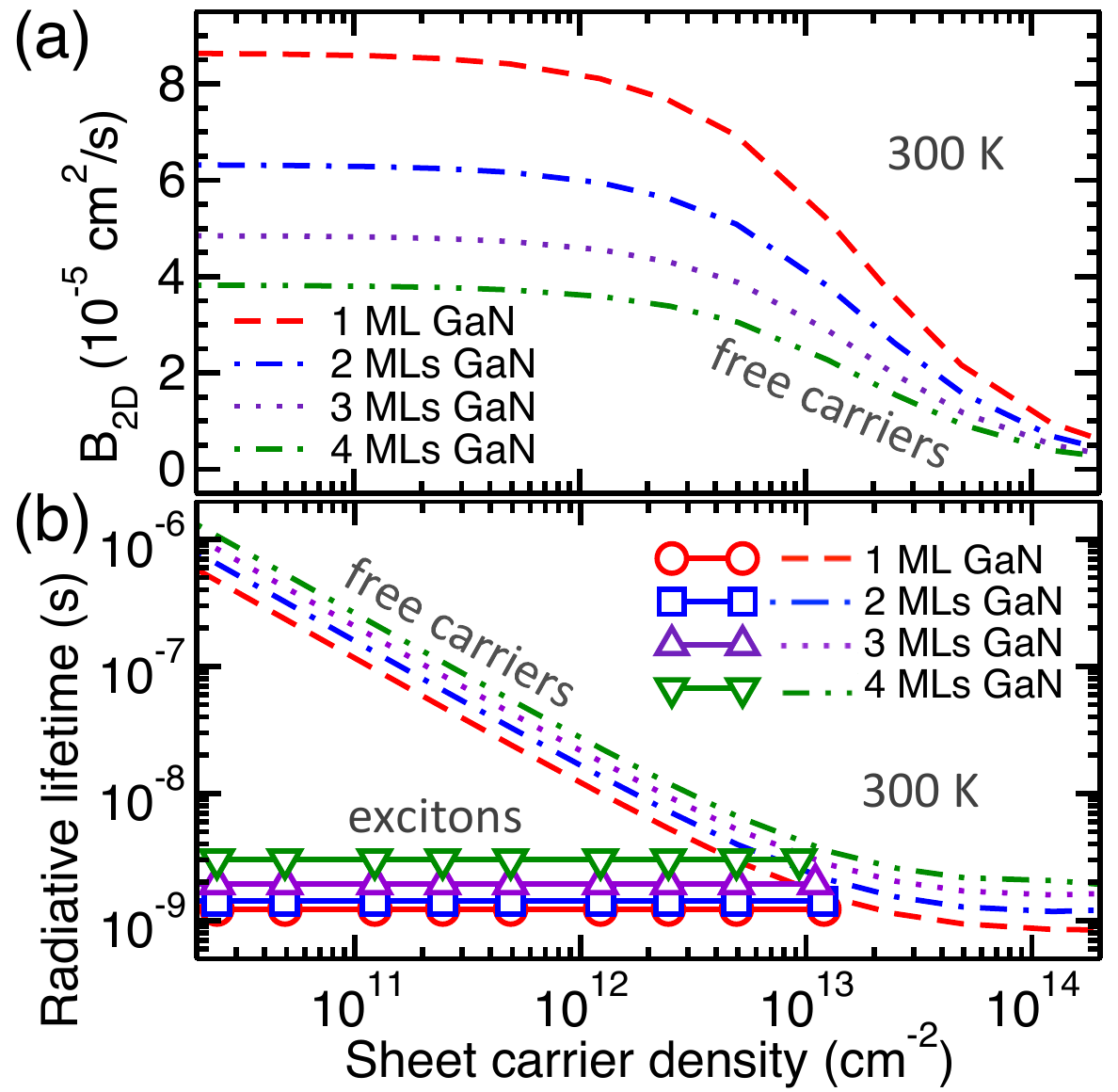}
\caption{\label{fig:radiative_lifetime}
(Color online) (a) Bimolecular radiative recombination coefficients ($B_{2D}$) and (b) radiative lifetimes of free carriers and excitons at 300 K in ML-thin GaN QWs as a function of sheet carrier density and well thickness. $B_{2D}$ decreases and the radiative lifetime increases as GaN well thickness increases due to the decreasing electron-hole overlap by the polarization field. Exciton radiative lifetimes are on the order of nanoseconds and up to several orders of magnitude shorter than those of free carriers. Excitons dissociate at densities above the Mott critical value at approximately 10$^{13}$ cm$^{-2}$.}
\end{figure}

To evaluate the radiative properties of atomically thin GaN QWs, we calculated the bimolecular radiative coefficient ($B$) of free carriers and the radiative recombination lifetimes of excitons (Fig. \ref{fig:radiative_lifetime}).
The $B$ coefficient determines the radiative recombination rate of free carriers per unit volume $R_{\text{rad}} = dn/dt$ as a function of carrier density $n$ according to $R_{\text{rad}} = Bn^2$.
Since the calculated $B$ coefficient depends on the simulation-cell size we compute the areal coefficients $B_{\text{2D}} = B/L$ ($L$ is the supercell period) as a function of sheet carrier density $n_{\text{2D}} = nL$, which are independent of simulation-cell size.
From the coefficients we also determine the free-carrier radiative lifetime by $\tau = 1/(B_{\text{2D}} n_{\text{2D}})$ (Fig. \ref{fig:radiative_lifetime}).
To elucidate the influence of confinement on radiative properties, we focus on the structures with the thickest AlN barrier for each GaN QW thickness (shown explicitly in Fig. \ref{fig:structure}).
The $B_{\text{2D}}$ coefficient increases and the radiative lifetime decreases with decreasing GaN QW thickness, demonstrating increasing radiative recombination rates by stronger quantum confinement (Fig. \ref{fig:radiative_lifetime}a), and the associated increased electron-hole wave function overlap $|F|^2$.
Moreover, at degenerate carrier densities (higher than $10^{13}$ cm$^{-2}$) the coefficient becomes a $1/n$ function of the density and the lifetime becomes constant due to phase-space filling.\cite{kioupakis2013temperature} with a value on the order of nanoseconds.

\begin{table*}
\centering
\caption{\label{tab:ultrathin_properties} 
Electronic, excitonic, and radiative recombination properties of atomically thin GaN QWs as a function of QW thickness: quasiparticle gap ($E_{\text{gap}}$), lowest singlet exciton energy ($E_S$) and binding energy ($\Delta E_{\text{b}}$),
energy splitting between the lowest singlet and lowest triplet excitons ($\Delta E_{ST}$), electron and hole envelope-function overlap integral ($|F|^2$),
exciton radiative lifetime at 0 K ($\tau_{\text{ex}}^0$),
exciton radiative lifetime at 300 K ($\tau_{_{\text{ex}}}$(300 K)),
hydrogenic model exciton lifetime at 300 K ($\tau_{_{\text{ex}}}^{\text{H}}$(300K)),
bimolecular recombination coefficient of free carriers in the limit of low carrier density at 300 K ($B_{\text{2D}}$(300 K)), 
exciton wave function at zero electron-hole separation ($|\psi(0)|^2$),
and Mott critical density for exciton dissociation ($n_{\text{crit}}$).
}
\begin{ruledtabular}
\begin{tabular}{cccccccccccc}





QW & $E_{\text{gap}}$ & $E_S$ & $\Delta E_{\text{b}}$ & $\Delta E_{ST}$ & $|F|^2$ &  $\tau_{\text{ex}}^0$ & $\tau_{_{\text{ex}}}$(300 K) & $\tau_{_{\text{ex}}}^{\text{H}}$(300 K) & $B_{\text{2D}}$(300 K) & $|\psi(0)|^2$ & $n_{\text{crit}}$ \\

 width & (eV) & (eV) & (meV) & (meV) &  & (ps) & (ns) & (ns) & ($10^{-5}$cm$^2$/s) & ($10^{12}$cm$^{-2}$) & ($10^{12}$cm$^{-2}$)\\
\hline
1 ML (0.26 nm) & 5.443 & 5.243 & 215 & 21 & 0.88 & 0.66 & 1.23 & 1.51 & 8.64 & 7.64 & 12.0 \\
2 MLs (0.52 nm) & 4.687 & 4.500 & 199 & 14 & 0.79 & 0.56 & 1.42 & 2.09 & 6.32 & 7.58 & 11.9 \\
3 MLs (0.77 nm) & 4.253 & 4.094 & 169 & 8 & 0.67 & 0.63 & 1.94 & 2.95 & 4.85 & 7.00 & 11.0 \\
4 MLs (1.03 nm) & 3.993 & 3.858 & 143 & 4 & 0.55 & 0.80 & 3.01 & 4.41 & 3.82 & 5.93 & 9.32 \\

\end{tabular}
\end{ruledtabular}
\end{table*}

\begin{figure}
\includegraphics[scale=0.6]{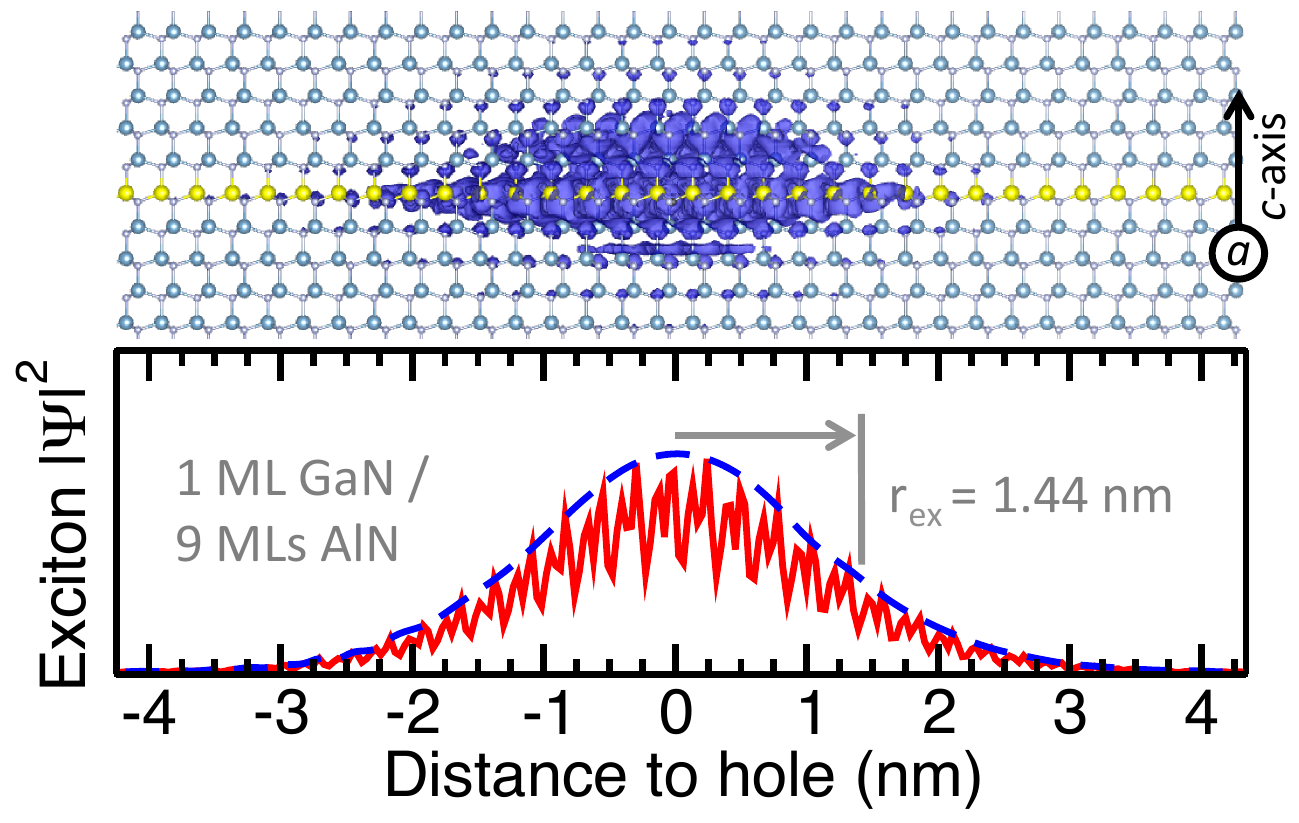}
\caption{\label{fig:exciton_wfns}
(Color online) The electron part of the squared exciton wave function in a 1 ML GaN QW. Top: three-dimensional isosurface plotted within the crystal structure.
Bottom: one-dimensional plot of the exciton wave function within the QW plane. The perpendicular directions have been integrated.
The exciton is strongly bound and highly localized (exciton radius of 1.44 nm at $e^{-1}\approx 0.368$ of $|\Psi|^2$ peak) due to the strong electron-hole interaction introduced by the extreme  confinement in the atomically thin well.
}
\end{figure}

Excitons, which we predict to be stable at room temperature in ML-thin GaN QWs, display fast radiative recombination at low density.
We calculated the radiative excitonic lifetimes from our BSE results and report the intrinsic bright exciton lifetime at zero temperature $\tau_{\text{ex}}^o$ as well as the thermally-averaged effective lifetime at 300 K $\tau_{\text{ex}}$(300 K) (Table \ref{tab:ultrathin_properties}).
The 0 K intrinsic lifetimes are approximately 0.6 ps, which are similar to calculated values in ML transition metal dichalcogenides\cite{palummo2015exciton}
The thermally averaged lifetimes at 300 K range from 1.23 to 3.01 ns, which are intermediate between calculated values for monolayers and bilayers of free-standing 2D GaN\cite{sanders2017electronic} and comparable to experimentally measured exciton lifetimes in 1.1 nm thick InGaN QWs (3--5 ns).\cite{langer2013room}
These exciton lifetimes of a few nanoseconds are several orders of magnitude shorter than free-carrier lifetimes at low carrier density (Fig. \ref{fig:radiative_lifetime}b) and independent of the free-carrier density due to the bound electron-hole pairs in excitons.\cite{langer2013room}
Carriers in bound exciton states experience a characteristic effective density equal to the exciton wave function evaluated at zero electron-hole separation, $|\psi(0)|^2$.\cite{hangleiter1993recombination}
To understand the origin of the exciton radiative lifetimes,
we evaluated the exciton wave function in atomically thin GaN QWs from our BSE results by inserting the $e^{-1}$ radius of the exciton envelope function, $r_{\text{ex}}$, (Fig.~\ref{fig:exciton_wfns}) into the ground-state solution of the two-dimensional hydrogen atom\cite{yang1991analytic} as $|\psi(0)|^2 = (2 \pi r_{\text{ex}}^2)^{-1}$.
The resulting effective densities $|\psi(0)|^2$ and hydrogenic-model radiative lifetimes, $\tau_{_{\text{ex}}}^{\text{H}}\text{(300 K)} = 1/(B_{\text{2D}}\text{(300 K)} |\psi(0)|^2)$, are given in Table \ref{tab:ultrathin_properties}. The model results are in excellent agreement with the values calculated from the explicit evaluation of excitonic lifetimes, $\tau_{_{\text{ex}}}$(300 K), and explain the origin of the nanosecond lifetimes.

While excitons are thermally stable if the characteristic thermal energy $k_bT$ does not exceed the binding energy, they dissociate into an electron-hole plasma if the carrier density exceeds the Mott critical value,\cite{gil2013iii} given in 2-dimensions by $n_{\text{crit}} = (2 r_{\text{ex}})^{-2}$.
The exciton Mott critical values in atomically thin GaN QWs are approximately $10^{13}$ cm$^{-2}$ (Table \ref{tab:ultrathin_properties}), which are lower than typical operating conditions in LEDs.

\color{black}
Our results assume ideal, abrupt interfaces. Electron microscopy experiments have demonstrated that the interfaces of short-period GaN/AlN superlattices can be atomically sharp.\cite{taniyasu2011polarization}
Recent work has also argued that the growth of ML GaN on AlN is self-limiting due to a balance between crystallization and evaporation of Ga adatoms.\cite{Kobayashi2019}
However, nonideal interfaces may occur during growth and can result in carrier localization within the plane, leading to broadening of photoemission spectra\cite{bayerl2016deep} or changing the exciton states from QW-like to quantum-dot-like\cite{verma2014tunnel}. Carrier localization may also affect the luminescence energy and the temperature dependence of the radiative lifetime. On the other hand, our main findings (that atomically thin wells increase the band gap on the scale of 1 eV and the exciton binding energy to hundreds of meV) are primarily controlled by the most confined direction (i.e., the ML thickness perpendicular to the layer) and are thus not expected to be drastically modified by fluctuations, as validated by the overall good agreement of our calculated luminescence energy with the experimental literature.
\color{black}

\color{black}
We also compare our calculated results for embedded GaN wells to earlier work for freestanding 2D GaN MLs. In comparison to the hydrogen-passivated tetrahedral  GaN,\cite{sanders2017electronic} embedded GaN MLs have a smaller band gap (5.443 eV versus 6.32 eV for freestanding\cite{sanders2017electronic}) due to reduced quantum confinement. Exciton binding is weaker for embedded (215 meV) than freestanding (1.31 eV)\cite{sanders2017electronic} MLs due to the screening of the Coulomb interaction by the AlN barriers, while the 300 K radiative lifetime is longer for embedded (1.23 ns) than freestanding (0.6 ns)\cite{sanders2017electronic} due to weaker exciton binding. In comparison, free-standing $sp^2$-bonded planar-hexagonal GaN\cite{peng2018room} has a smaller calculated gap (4.382--4.1 eV\cite{peng2018room,prete2017tunable}) due to the qualitatively different character of its band-edge states compared to the $sp^3$-bonded tetrahedral GaN MLs.
\color{black}

Atomically thin GaN QWs are promising for highly efficient deep-UV optoelectronics.
Extreme confinement in ML-thin GaN wells increases the gap into the deep UV, as well as the spatial overlap of electrons and holes.
Confinement preserves the TE-polarized emission of GaN, which facilitates light extraction in c-plane devices and resolves one of the problems of high-aluminum-content AlGaN.\cite{taniyasu2011polarization,islam2017mbe}
However, the advantages of ML confinement are more evident when considering the role of excitons,
which are stable against thermal dissociation at room temperature and dominate carrier recombination. 
Excitonic effects simultaneously increase the radiative and the Auger recombination rates.\cite{langer2013room} However, the simultaneous increase of the two rates by similar amounts is beneficial for the high-power LED efficiency since faster recombination reduces the steady-state carrier density and thus the fraction of carriers that recombine via the Auger process.
We also anticipate that the Auger coefficient of GaN MLs is smaller than AlGaN alloys because alloy-assisted indirect Auger, which is important in InGaN,\cite{Kioupakis2015} is absent in binary GaN and may underlie the high efficiencies of few-ML GaN LEDs.\cite{Rong2016,islam2017deep,Kobayashi2019}
We therefore expect that the increased recombination rates by exciton stabilization may increase the efficiency of ML-GaN-based deep-UV LEDs compared to AlGaN alloys.

In conclusion, we investigated the electronic, excitonic, and radiative properties of  atomically thin GaN/AlN QWs from first-principles calculations.
We demonstrate that extreme quantum confinement shifts the band gap of GaN into the deep-UV and increases the exciton binding energy up to 215 meV, stabilizing excitons against thermal dissociation at room temperature.
Our results indicate that atomically thin GaN QWs can achieve TE-polarized deep-UV light emission with potentially higher internal quantum efficiency than AlGaN QW devices.

See supplementary material for \color{black}compuational details and \color{black} band structures of 1-4 ML GaN structures (Fig.~\ref{fig:structure}).

This work was supported by the NSF DMREF program (1534221).
Computational resources were provided by the DOE NERSC facility (DE-AC02-05CH11231).

%

\end{document}


\title{Supplementary Material: Room-temperature stability of excitons and transverse-electric polarized deep-ultraviolet luminescence in atomically thin GaN quantum wells}

\author{Dylan Bayerl}
\affiliation{Department of Materials Science and Engineering, University of Michigan, Ann Arbor, MI 48109, USA}
\author{Emmanouil Kioupakis}
\email{kioup@umich.edu}
\affiliation{Department of Materials Science and Engineering, University of Michigan, Ann Arbor, MI 48109, USA}

\date{\today}

\maketitle

\section{Details of first-principles calculations}

Our first-principles methodology employs density functional theory (DFT) and many-body perturbation theory, including the GW and Bethe-Salpeter equation (BSE) methods.
We performed DFT calculations within the local density approximation (LDA) with the plane-wave norm-conserving pseudopotential method.\cite{quantumespresso} 
Electronic and optical properties were calculated using a semicore Ga pseudopotential including 3$s$ and 3$p$ orbitals in the valence.
Quasiparticle corrections to the band gap were calculated with the GW method\cite{berkeleyGW} using the static-remainder\cite{statrem_deslippe} correction and the generalized plasmon-pole model of Hybertsen and Louie,\cite{GW_hybertsenlouie} excluding the Ga semicore shell from the plasmon-pole sum rule.\cite{malone2013quasiparticle}
Optical properties were calculated by solving the BSE to obtain the exciton energy spectrum,\cite{rohlfing2000electron,berkeleyGW} yielding exciton binding energies and optical gaps.
K-point grids ranging from $8\times 8 \times 4$ to $8\times 8 \times 1$ were used to preserve uniform sampling of the Brillouin zone.
The plane-wave cutoff and screening cutoff energies used in DFT and GW calculations were 250 Ry and 34 Ry, respectively.
The number of bands used in the Coulomb-hole summations included states with eigenenergies up to 50\% of the screening cutoff energy, which converged the quasiparticle gap throughout the Brillouin zone to within 50 meV of that of the extrapolated infinite-band summation. 

Simulated crystal structures were pseudomorphically strained to the AlN barrier with the basal plane lattice constant fixed to the experimental value of bulk AlN (0.3112 nm)\cite{levinshtein2001properties} and the c-axis unconstrained to minimize strain energy and mimic pseudomorphic growth on an AlN buffer. 
The relaxed c-axis of the AlN barrier varies with barrier thickness and distance from the GaN layer due to the lattice mismatch, but was on average 0.8\% larger than the experimental value (0.4982 nm). 
Minimization of strain energy resulted in extension of the GaN layer c-axis by 1.5\% to 2.0\% relative to the experimental value of bulk GaN (0.5178 nm).

\begin{figure}
\includegraphics[width=\columnwidth]{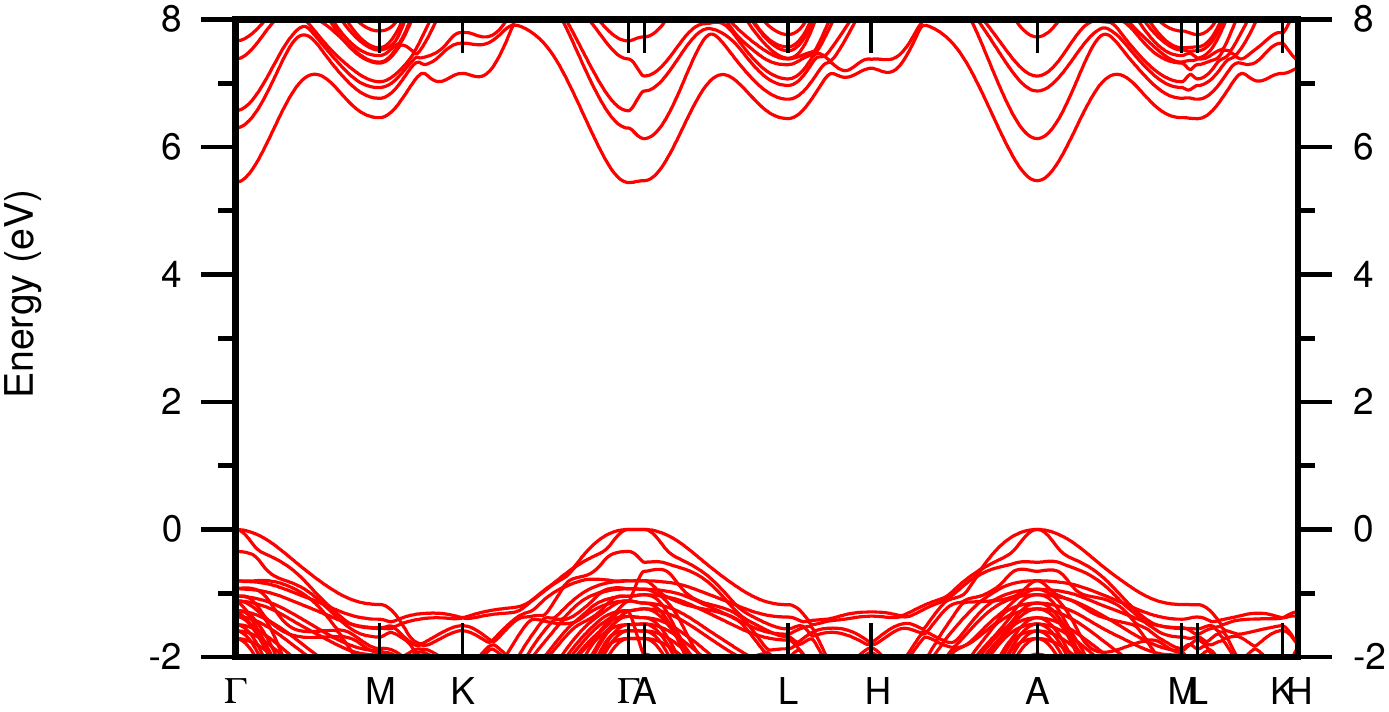}
\caption{\label{fig:u5g1}
(Color online) Band structure of a superlattice consisting of a GaN quantum well with a thickness of 1 monolayer and an AlN barrier with a thickness of 9 monolayers.  
}
\end{figure}

\begin{figure}
\includegraphics[width=\columnwidth]{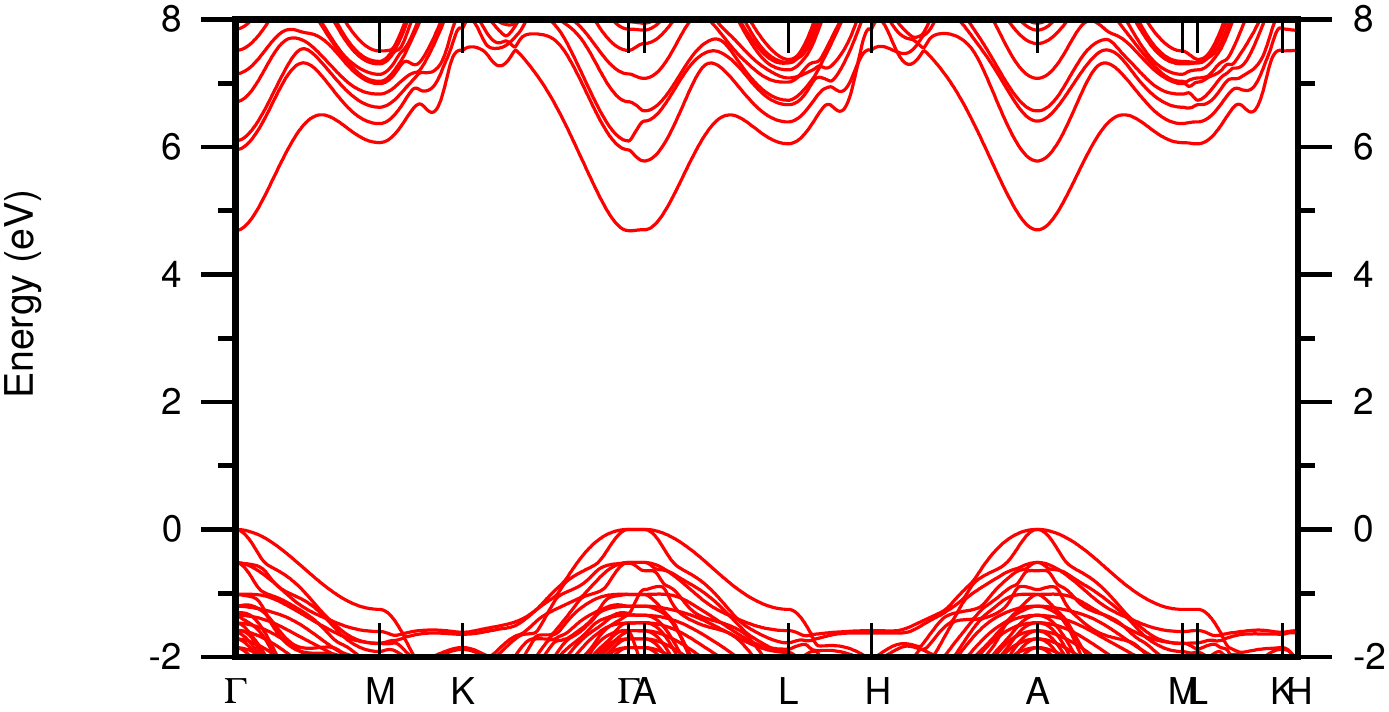}
\caption{\label{fig:u5g1}
(Color online) Band structure of a superlattice consisting of a GaN quantum well with a thickness of 2 monolayers and an AlN barrier with a thickness of 8 monolayers.  
}
\end{figure}

\begin{figure}
\includegraphics[width=\columnwidth]{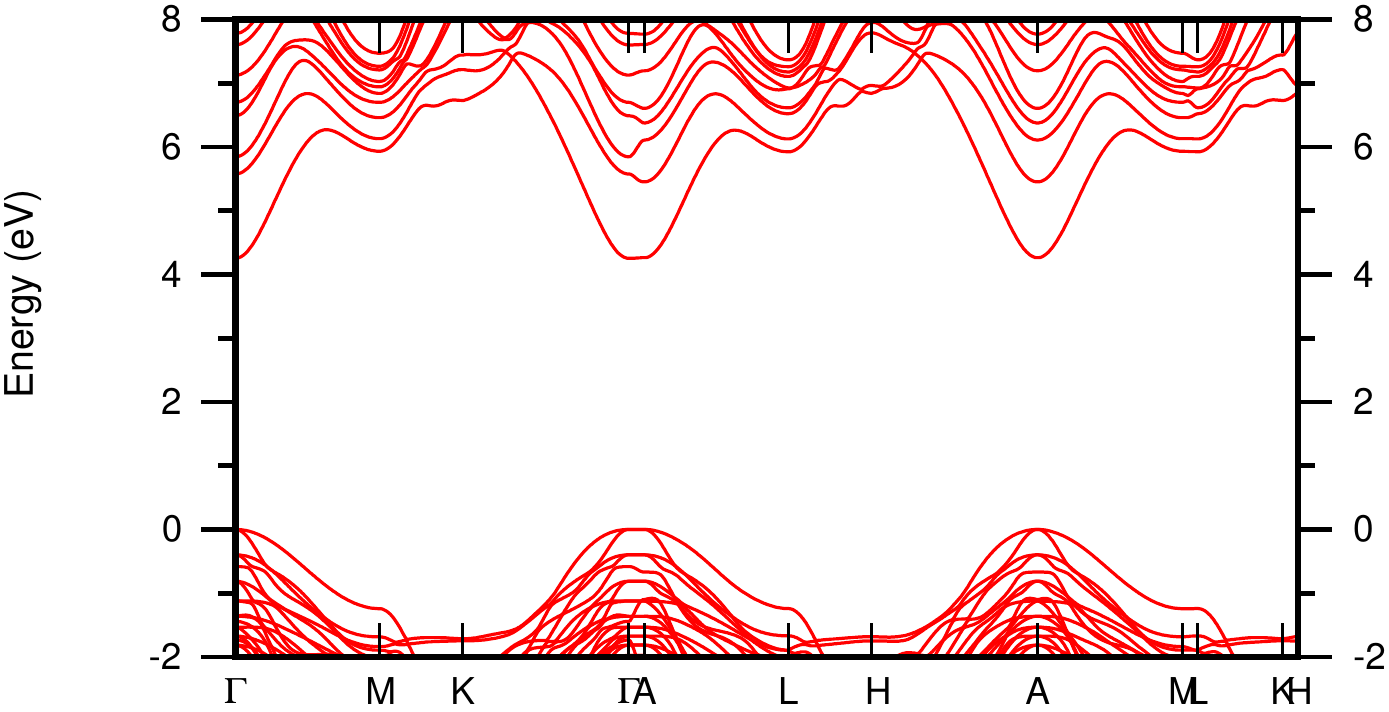}
\caption{\label{fig:u5g1}
(Color online) Band structure of a superlattice consisting of a GaN quantum well with a thickness of 3 monolayers and an AlN barrier with a thickness of 7 monolayers.  
}
\end{figure}

\begin{figure}
\includegraphics[width=\columnwidth]{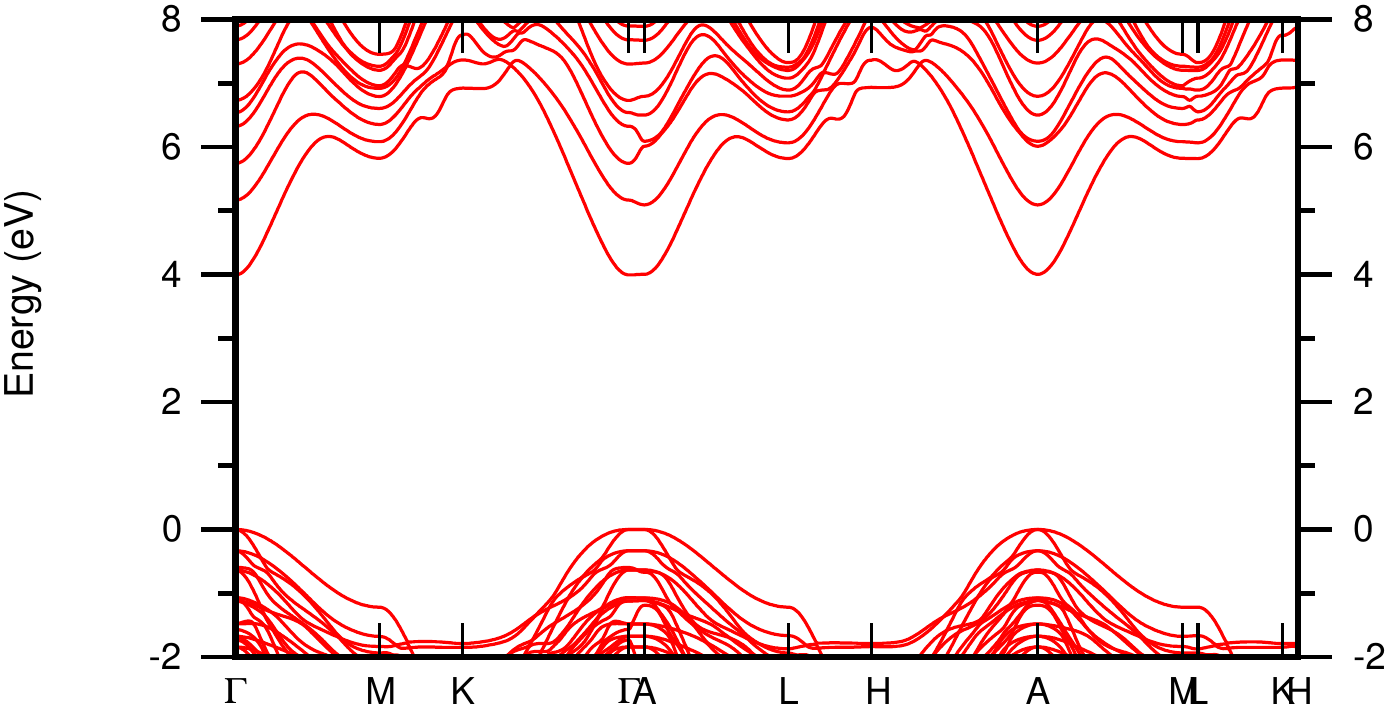}
\caption{\label{fig:u5g1}
(Color online) Band structure of a superlattice consisting of a GaN quantum well with a thickness of 4 monolayers and an AlN barrier with a thickness of 6 monolayers.  
}
\end{figure}

\clearpage
\bibliography{algan_qwsl}